\documentclass[aps,pra,twocolumn,amsmath,amssymb,nofootinbib,superscriptaddress]{revtex4-1}

\newcommand{\bra}[1]{\left\langle#1\right\rvert}
\newcommand{\ket}[1]{\left\lvert#1\right\rangle}
\newcommand{\ketbra}[2]{\left\lvert{#1}\middle\rangle\!\middle\langle{#2}\right\rvert}

\newcommand{\expect}[1]{\left\langle{#1}\right\rangle}
\newcommand{\tr}{\mathrm{Tr}}
\newcommand{\ptr}[2]{\mathrm{Tr}_{#1}\left( #2 \right)}
\newcommand{\set}[1]{\left\lbrace #1 \right\rbrace}
\newcommand{\sgn}[0]{\text{sgn}}

\usepackage[colorlinks,linkcolor=blue,citecolor=blue,allcolors=blue]{hyperref} %links for citations
\usepackage[pdftex]{graphicx}
\usepackage{subfigure}
\usepackage{mathrsfs}
\usepackage{subfigure}
\usepackage{float}
\usepackage{bm}
\usepackage{soul}
\usepackage{color}
\usepackage[usenames,dvipsnames]{xcolor}
\usepackage{framed}

\definecolor{celadon}{rgb}{0.67, 0.88, 0.69}

\definecolor{shadecolor}{rgb}{1.0, 0.99, 0.82}

\begin{document}

\title{Emergence of Opposing Arrows of Time in Open Quantum Systems}

\author{Thomas Guff}
\email[]{t.guff@surrey.ac.uk}
\affiliation{School of Mathematics and Physics, University of Surrey, GU2 7XH Guildford, United Kingdom}

\author{Chintalpati Umashankar Shastry}
\affiliation{School of Mathematics and Physics, University of Surrey, GU2 7XH Guildford, United Kingdom}
\affiliation{Leverhulme Quantum Biology Doctoral Training Centre, University of Surrey, GU2 7XH Guildford, United Kingdom}

\author{Andrea Rocco}
\email[]{a.rocco@surrey.ac.uk}
\affiliation{School of Mathematics and Physics, University of Surrey, GU2 7XH Guildford, United Kingdom}
\affiliation{School of Biosciences, University of Surrey, GU2 7XH Guildford, United Kingdom}

\date{\today}

\begin{abstract}

Deriving an arrow of time from time-reversal symmetric microscopic dynamics is a fundamental open problem in many areas of physics, ranging from cosmology, to particle physics, to thermodynamics and statistical mechanics. 
Here we focus on the derivation of the arrow of time in open quantum systems and study precisely how time-reversal symmetry is broken. 
This derivation involves the Markov approximation applied to a system interacting with an infinite heat bath. We find that the Markov approximation does not imply a violation of time-reversal symmetry. Our results show instead that the time-reversal symmetry is maintained in the derived equations of motion. 
This imposes a time-symmetric formulation of quantum Brownian motion, Lindblad and Pauli master equations, which hence describe thermalisation that may occur into two opposing time directions. 
As a consequence, we argue that these dynamics are better described by a time-symmetric definition of Markovianity. 
Our results may reflect on the formulations of the arrow of time in thermodynamics, cosmology, and quantum mechanics. 

\end{abstract}

\maketitle

\section{Introduction}\label{sec:intro}

The existence of an `arrow of time' \cite{Eddington28} undeniably characterizes our common experience. 
We keep memory of the past, not only as living beings, but also through objective evidence, such as pictures and fossils. 
In contrast, we do not have memory of the future. The arrow of time describes the clear asymmetry that makes the past intrinsically different from the future.

And yet, the fundamental laws of physics in both the classical and the quantum realms do not manifest any intrinsic arrow of time.  Newton's equations are time-reversal symmetric, as well as Schr{\"o}dinger's equation. 
As a consequence, backward-in-time motion is equally possible as forward-in-time motion. The motion of planets around the Sun, as much as the oscillations of a pendulum, are phenomena that do not distinguish the past from the future. 
A movie of these processes, played backwards, would still represent a perfectly legitimate physical phenomenon.

As a result, the conundrum arises in many areas of physics of how to derive an arrow of time from underlying time-symmetric dynamics, both classically \cite{Roduner22} and quantum-mechanically \cite{Strasberg21}.

This puzzling issue was at the heart of the debate taking place at the end of the 19th century between Boltzmann \cite{Boltzmann1872} on the one side and Loschmidt \cite{Bader01} and Zermelo \cite{Steckline83} on the other. Boltzmann’s celebrated $H$-theorem is the best known attempt to derive classically an arrow of time -- so-called `thermodynamic arrow of time' -- using probabilistic arguments and phase space coarse-graining. 
However, Loschmidt and Zermelo saw a contradiction between Boltzmann’s conclusions and the underlying Hamiltonian time-symmetric mechanics. The former argued that time-reversal symmetry implies that for every trajectory of a system there exists a time-reversed trajectory for which $H$ increases \cite{Bader01}; and the latter argued that any bounded system should display Poincar{\'e} recurrences, where the system returns after some time to the initial state if discrete, or arbitrarily close to it if continuous \cite{Steckline83}. 
Boltzmann's counter arguments to these objections were both probabilistic and practical, with any violation of the second law deemed either extremely unlikely (the fluctuations leading to large deviations from the expected minimum of $H$ would be exceedingly rare) or not observable (Poincar{\'e} recurrences would be expected over a time larger than the age of the universe).

A similar debate takes place nowadays in cosmology, where the derivation of a `cosmological arrow of time' is thought to be related to the expansion of the universe from the Big Bang under the assumption of an initial low-entropy state (the so-called `past hypothesis' \cite{Albert00}), and is believed to emerge from inflationary models of the universe \cite{Carroll05}. However, fundamental questions have started to emerge in this field as well on the very uniqueness of this arrow of time in gravitational models of $N$-body systems \cite{Barbour14}, reflecting, once again, the intimate problematics between time-reversal symmetry of Hamiltonian systems and the emergence of irreversibility.

Of more relevance in this paper is the so-called `quantum arrow of time', which can be associated to the relaxational and decoherence processes, responsible for the irreversible loss of information of a quantum system becoming increasingly entangled with its environment \cite{Linden09}, or because of quantum measurements \cite{Rubino21}.

In the context of open quantum systems, several approaches {are currently considered} to tackle the derivation of the quantum arrow of time. 
These range from redefining the statistical properties of ensembles from first principles \cite{Popescu06}, to adopting  information-theoretical frameworks \cite{Goold16}, or constructor theory \cite{Deutsch13,Marletto22}. In the same spirit, non-negative entropy production rates have been obtained explicitly in reduced dynamics of open quantum systems \cite{Deffner11,Santos17,Batalhao19,VanVu21}. In this paper too, we adopt the framework of open quantum systems and study the arrow of time through the time-reversal symmetry of the ensuing quantum equations of motion.

An open quantum system consists of a system of interest interacting with its environment \cite{Lindenberg90}. 
This combined system is assumed to be described by a global Hamiltonian, which is time-reversal symmetric. 
A standard reduction method is then chosen to trace over the degrees of freedom of the environment \cite{Nakajima58, Zwanzig60, Mori65}. 
This leads typically to non-Markovian dynamics, wherein the reduced equations of motion contain the entire history of the system, mathematically described by convolution integrals and memory kernels \cite{Hu92,Breuer07,Lally22}. 

If the memory kernels of the system decay quickly enough, the system can be approximated as Markovian at sufficiently long times.
%A Markovian system cannot recover any information or energy lost to the environment, and the reduced dynamics describe dissipative evolution into the future, towards a stationary state.
Markovian descriptions of the system of interest are usually cast in terms of the quantum Langevin equation \cite{Gardiner2004}, and Lindblad master equations \cite{Lindblad76,Gorini76}.
Since these equations typically break time-reversal symmetry, the implication is usually made that a clear arrow of time has been derived. 

Here we will show that this derivation produces in fact two arrows of time, not one, and in so doing preserves time-reversal symmetry in the reduced Markovian dynamics of the system.
We will do this by considering several well-known derivations of Markovian dynamics in open quantum systems. In particular we study the model of a system  coupled to an environment consisting of a large number of harmonic oscillators. 
This model has been introduced in the pioneering path integral formulation of Caldeira and Leggett \cite{Feynman63,Caldeira1983}, and is broadly recognized as the prototypical model of an open quantum system \cite{Weiss2012}.

We find that not only Markovian time-reversal symmetry persists in the quantum Langevin equation (previously noted in \cite{Hanggi2005}), but also in the Brownian motion master equation. Furthermore, we also revisit the standard microscopic derivations of the Lindblad master equation and the Pauli master equation, and find also in these cases that the time-reversal symmetry is maintained.

The Markov approximation only involves  assumptions on the structure of the bath, and the interaction with the system. 
For this reason it has to be implemented in an agnostic way with respect to the arrow of time. 
As a consequence, the {subsequent equations of motion predict that the system will approach equilibrium both forwards and backwards in time, i.e. the system is dissipative and decohering in both temporal directions. 
In contrast, the violation of time-reversal symmetry in the standard derivation of quantum dissipative dynamics happens only when the Markov approximation is imposed asymmetrically by hand. 

Our results show that time-reversal symmetry is not inherently at odds with dissipation and decoherence.
Dissipative dynamics, and the second law, still hold in our derivation, once the arrow of time has been chosen \emph{a priori}. 
The time-reversal symmetry of the full system is preserved in the reduced dynamics because two opposing choices of the arrow of time 	 are in fact possible.
We also argue that it is not time-reversal symmetry, but time \emph{translation} symmetry that is broken by the Markov approximation.

Our results do not contradict the macroscopic description of the second law, whereby the system tends to evolve, with overwhelming likelihood, to macrostates associated with the largest number of microstates.
Rather, we are showing that the \emph{microscopic} derivations of an arrow of time using employing the Markov approximation to open systems does not in fact break time-reversal symmetry.

The structure of this paper is as follows. 
In Sec. \ref{sec:markov} we present some preliminary considerations and definitions on the Markov approximation and the time-reversal transformation.
In Sec. \ref{redq} we study a general derivation of Markovian dynamics in models of open quantum systems, and show that we consistently obtain time-reversal symmetric quantum Langevin equations and Brownian motion master equations. 
Secs. \ref{sec:Lindblad} and \ref{sec:Pauli} generalize our approach to the derivation of the Lindblad and the Pauli master equations respectively, while in Sec. \ref{timetransl} we discuss the breakdown of time-translation symmetry caused by the Markov approximation. 
Finally, in Sec. \ref{concl} we provide some concluding remarks.

\section{The time-symmetric Markov approximation\label{sec:markov}}

Markov processes are stochastic processes with no memory. That is, processes for which the probability of a future event depends only on the current state of the system and not on its past.

Classically, a stochastic process described by the probability distribution $\mathbb{P}(x,t)$ for the random variable $X$ is \emph{Markovian} if for any finite set of successive times $t_{1} < t_{2} < \dots < t_{n}$, and for all possible values $\set{x_{1},\dots,x_{n}}$ of the random variable $X$, the following relation for conditional probabilities holds: 
\begin{equation}
\mathbb{P}(x_{n},t_{n} |  x_{1},t_{1};\dots;x_{n-1},t_{n-1}) = \mathbb{P}(x_{n},t_{n} | x_{n-1},t_{n-1}).\label{eq:markovian}
\end{equation}

In quantum mechanics the probabilistic evolution of a system degree of freedom is encoded into the density operator $\hat{\rho}(t)$ for that system. 
For a system uncorrelated from its environment at time $t=0$, the evolution of the system for time $t>0$ is described by a completely positive trace-preserving (CPTP) map $\hat{\mathcal{E}}(t)$,
\begin{equation}
\hat{\mathcal{E}}(t)(\hat{\rho}(0)) = \hat{\rho}(t). 
\end{equation}
The evolution map $\hat{\mathcal{E}}(t)$ is \emph{Markovian} if for all times $t_{1}, t_{2}>0$,
\begin{equation}
\hat{\mathcal{E}}(t_{1}+t_{2}) = \hat{\mathcal{E}}(t_{1})\hat{\mathcal{E}}(t_{2}), \label{eq:divisible}
\end{equation}
or in other words $\set{\hat{\mathcal{E}}(t)_{t>0}}$ is a quantum dynamical semigroup \cite{BreuerPetruccione07}.

An example of quantum Markovian evolution is given by the Lindblad equation \cite{Lindblad76,Gorini76},
\begin{equation}
\frac{d\hat{\rho}}{dt}(t) = \hat{\mathcal{L}}\hat{\rho}(t), \label{eq:Lindblad}
\end{equation}
where the resulting evolution map $\hat{\mathcal{E}}(t) = \exp(\hat{\mathcal{L}} t)$ is CPTP and satisfies \eqref{eq:divisible}. This equation can model relaxation behaviour: if the eigenvalues of the Lindbladian $\hat{\mathcal{L}}$ have a non-zero real part, then the off-diagonal terms of $\hat{\rho}(t)$ will exponentially decay to zero in the appropriate basis \cite{BreuerPetruccione07}.

Equation (\ref{eq:Lindblad}) is typically not time-reversal symmetric. In this paper we will examine several derivations of equations such as this from time-reversal symmetric Hamiltonian models of a system weakly interacting with an environment. In all cases we find that the system does not satisfy precisely these equations but time-reversal symmetric versions of them. In fact equations such as \eqref{eq:Lindblad} arise when one ignores the past evolution of the system.

We note here that both classical and quantum definitions of Markovian dynamics contain an explicit arrow of time, in that they are defined for positive times. This is consistent in the quantum case with the solutions of \eqref{eq:Lindblad} only being defined for $t>0$. In particular, if the Lindblad equation (\ref{eq:Lindblad}) predicts exponential decay for off-diagonal terms of the density matrix, extrapolating backwards for sufficiently long times will eventually produce a state which is no longer positive semi-definite, since the off-diagonal terms are bounded by the diagonal terms to retain positivity. This is in contrast with typical solutions of Schr\"odinger's equation which give trajectories for all $t$.

Therefore Markovian dynamics are typically only valid after a particular reference time, commonly chosen to be $t=0$. This introduces a clear distinction between the future and the past, and a possible origin for the violation of time-reversal symmetry.

This observation agrees with the Markov condition being only defined for an increasing collection of timestamps as in the standard definitions (\ref{eq:markovian}) and (\ref{eq:divisible}). The requirement that Markovian dynamics be well defined also at negative times suggests an extension of the Markov property so that it holds symmetrically into two time directions. 
More specifically, we propose that a classical stochastic process $\mathbb{P}(X,t)$ should be considered Markovian if \eqref{eq:markovian} holds for any finite set of times $0<t_{1}<t_{2}<\dots<t_{n}$ or $0>t_{1}>t_{2}>\dots>t_{n}$ and for all possible values $\set{x_{1},\dots,x_{n}}$ of the random variable $X$. 

Likewise, we will say that a quantum evolution map $\hat{\mathcal{E}}(t)$ is Markovian if the semigroup property \eqref{eq:divisible} holds for all times $t_{1},t_{2}>0$ or $t_{1},t_{2}<0$.
This distinction between times being entirely positive or entirely negative is necessary to avoid triviality; since, for example, for quantum systems a CPTP map can be reversed by another CPTP map if and only if the dynamics are unitary \cite{Rivas2012}. 

In the rest of this paper we show that this symmetric definition in fact arises naturally in standard derivations of quantum reduced Markovian dynamics. As a consequence we demonstrate that the Markov approximation is entirely compatible with time-reversal symmetry, and that the apparent violation of it in Markovian structures like (\ref{eq:Lindblad}) in fact occurs by actively ignoring the dynamics for $t<0$.
{In section \ref{sec:CL} we will derive a time-reversal symmetric version of the Brownian motion master equation.
While it satisfies the divisibility criterion \eqref{eq:divisible} symmetrically, the standard application of the Markov approximation breaks the complete positivity of the evolution map.
Given its ubiquitous use as the quantum description of Brownian motion, it is still an insightful example of how the Markov approximation retains time-reversal symmetry.

\section{Evolution of Open Quantum Systems} \label{redq}

One of the most common models of an open quantum system adopted to derive Markovian dynamics consists of a system of interest linearly coupled to a large environment composed of harmonic oscillators \cite{Ford65,Cortes85,Feynman63,Caldeira1983}.
By choosing a suitable spectral density it can be shown that the system follows a quantum Langevin equation, from which it is possible to derive the quantum Brownian motion master equation by tracing over the environment. Here we study this derivation and pay particular attention to the consequences of the Markov approximation on time-reversal symmetry. In the subsequent sections we also analyse the Lindblad and Pauli master equations. In all cases, we find that the Markov approximation does not violate time-reversal symmetry.

\subsection{The model}

The model consists of a particle of mass $M$ characterized by a single positional degree of freedom $\hat{Q}$ and corresponding momentum $\hat{P}$, in a time-independent potential $V(\hat{Q})$. The environment is modelled using $N$ harmonic oscillators, each with position coordinates $\hat{q}_{k}$, momenta $\hat{p}_{k}$, frequency $\omega_{k}$ and mass $m_{k}$, with $k=1,\dots, N$. Each oscillator is coupled to the system of interest through its respective position coordinates, and no coupling is assumed between oscillators.

The Hamiltonian for system and environment is
\begin{equation}
\hat{H} = \hat{H}_{\textsc{S}} + \hat{H}_{\text{B}} + \hat{H}_{\textsc{SB}} \label{eq:H}
\end{equation}
where
\begin{subequations}
\begin{align}
&\hat{H}_{\textsc{S}} = \frac{\hat{P}^{2}}{2M} + V(\hat{Q}), \label{at}\\
&\hat{H}_{\textsc{B}}^{(k)} = \sum_{k=1}^{N}\left(\frac{\hat{p}_{k}^{2}}{2m_{k}} + \frac{m_{k}\omega_{k}^{2}\hat{q}_{k}^{2}}{2}\right), \label{bt}\\
&\hat{H}_{\textsc{SB}}^{(k)} = \sum_{k=1}^{N}\left(g_{k} \hat{q}_{k} \hat{Q} + \frac{g_{k}^{2}}{2m_{k}\omega_{k}^{2}}\hat{Q}^{2}\right). \label{ct}
\end{align}
\end{subequations}
Here $g_{k}$ is the coupling constant between the $k^{\text{th}}$ oscillator and the system, and the second term in Eq. (\ref{ct}) is inserted for convenience so as to cancel the environment-induced renormalization of the potential $V(\hat{Q})$ \cite{Lindenberg90}. This model can be recast in the path integral formalism introduced by Feynman and Vernon \cite{Feynman63} and further developed by Caldeira and Leggett \cite{Caldeira1983}. However, we adopt here the Heisenberg picture method introduced in \cite{Gardiner2004}, which will allow us to derive directly the corresponding quantum Langevin equation.

By following \cite{Gardiner2004}, the equations of motion for the system of interest can be written as
\begin{equation}
\frac{d^2\hat{Q}}{dt^2}(t) + V^{\prime}(\hat{Q}(t))+\frac{1}{M}\int_{0}^{t} k(t-t^{\prime})\hat{P}(t^{\prime})\, dt^{\prime}+ k(t)\hat{Q}(0) = \hat{f}(t)   \label{eq:GLE},  
\end{equation}
where
\begin{equation}
k(t) = \sum_{i=1}^{N} \frac{g_{k}^{2}}{m_{k}\omega_{k}^{2}}\cos(\omega_{k} t), \label{eq:kernel}\\
\end{equation}
and
\begin{equation}
\hat{f}(t) = \sum_{i=1}^{N} \left( g_{k}\hat{q}_{k}(0)\cos(\omega_{k}t)+\frac{g_{k}\hat{p}_{k}(0)}{m_{k}\omega_{k}}\sin(\omega_{k}t)\right),\label{eq:stochf}
\end{equation}
are known as the memory kernel and the stochastic force respectively.

Equation \eqref{eq:GLE} is a deterministic equation for $\hat{Q}(t)$. However, if the initial condition of the bath is a thermal state $\rho_{\text{th}}$ at temperature $T$ with respect to the bath Hamiltonian $\hat{H}_{\textsc{B}}$ then the average force term is zero,
\begin{equation}
\expect{\hat{f}(t)} = \tr(\hat{f}(t)\rho_{\text{th}}) = 0, \label{eq:zeroavg}
\end{equation}
and the autocorrelation function in the quantum case has the form 
\begin{equation}
\expect{\set{\hat{f}(t),\hat{f}(t^{\prime})}} = \sum_{k=1}^{N} \frac{\hbar g_{k}^{2}} {m_{k}\omega_{k}} \coth\left(\frac{\hbar \omega_{k}}{2 k_{B}T}\right)\cos(\omega_{k}(t-t^{\prime})),\label{eq:qcorr}
\end{equation}
where $\set{\cdot, \cdot}$ is the anti-commutator. 

Eq. \eqref{eq:GLE} is then referred to as the \emph{generalised quantum Langevin equation}; generalised in that the equation of motion doesn't simply have a term proportional to the velocity, but rather one which is the convolution of the velocity with the memory kernel $k(t)$. It is clearly not Markovian, as it features an integral over the history of the velocity of the particle from time $t=0$.

\subsection{Time-Reversal Transformation} \label{thetasec}

Time-reversal symmetry (Fig.~\ref{TRFig}) captures the notion that by watching the motion of a system one cannot tell whether it is moving forwards or backwards, since both the forward and backwards trajectories satisfy the same equation of motion. A classic example of a time-reversal symmetric system is a planet orbiting a star. By contrast, an ice cube melting into liquid water is not time-reversal
symmetric, since the backward motion is never seen.

It is important at this point to distinguish between several phrases often conflated or used interchangeably, which we want to keep separate: invertibility, reversibility, and time-reversal symmetry.
\begin{enumerate}
\item \emph{Invertibility:} Open quantum dynamics described by a CPTP map $\hat{\mathcal{E}}$ are said to be \emph{invertible} if there exists an inverse map $\hat{\mathcal{E}}^{-1}$ such that $\hat{\mathcal{E}}^{-1}\hat{\mathcal{E}} = \hat{\mathbb{I}}$.
Under certain conditions invertibility is equivalent to divisibility of a universal dynamical map \cite{Jeknic-Dugic2023}.
\item \emph{Reversibility:} Open quantum dynamics are said to be reversible if they are invertible, and furthermore if the inverse map $\hat{\mathcal{E}}^{-1}$ is a CPTP map.
This is only possible if the dynamics are unitary \cite{Rivas2012}.
\item \emph{Time-Reversal Symmetry:} Open quantum dynamics are said to be time-reversal symmetric if for every solution of the equations of motion, the time-reversed trajectory is also a solution.
\end{enumerate}

As we have defined them, these concepts are not interchangeable. 
Clearly, reversibility implies invertibility but invertibility does not imply reversiblilty.
A system can be time-reversal symmetric even though it is not reversible (as we will show in this paper).
A system can be reversible (unitary) but not time-reversal symmetric.
Consider the simple quantum harmonic oscillator, and transform the Hamiltonian as $\hat{H} \to e^{-\mu t} \hat{H}$.
The corresponding time-dependent Schrödinger equation will remain unitary but no longer time-reversal symmetric.
Furthermore, as we show in this section, the generalised Langevin equation is time-reversal symmetric, but in general it is highly non-Markovian, and therefore non-invertible \cite{Jeknic-Dugic2023}.

In quantum mechanics, the time-reversal transformation is implemented using the time-reversal operator $\hat{\Theta}$, defined as the anti-unitary operator \cite{Wigner1959}, which reverses the momentum of quantum particles, without affecting their position, while maintaining their canonical commutation relations.
\begin{equation}
\hat{\Theta} \hat{\bm{q}} \hat{\Theta}^{-1} = \hat{\bm{q}}, \quad \hat{\Theta} \hat{\bm{p}} \hat{\Theta}^{-1} = -\hat{\bm{p}}, 
\end{equation}
where $\hat{\bm{q}}=\left(\hat{Q},\hat{q}_1,...,\hat{q}_N\right)$ and $\hat{\bm{p}}=\left(\hat{P},\hat{p}_1,...,\hat{p}_N\right)$ are the coordinates and the momenta respectively of all particles in the total system, and the time-reversal operator acts element-wise on $\bm{\hat{q}}$ and $\bm{\hat{p}}$.
\begin{figure}[t]
\centering
\includegraphics[scale=0.55]{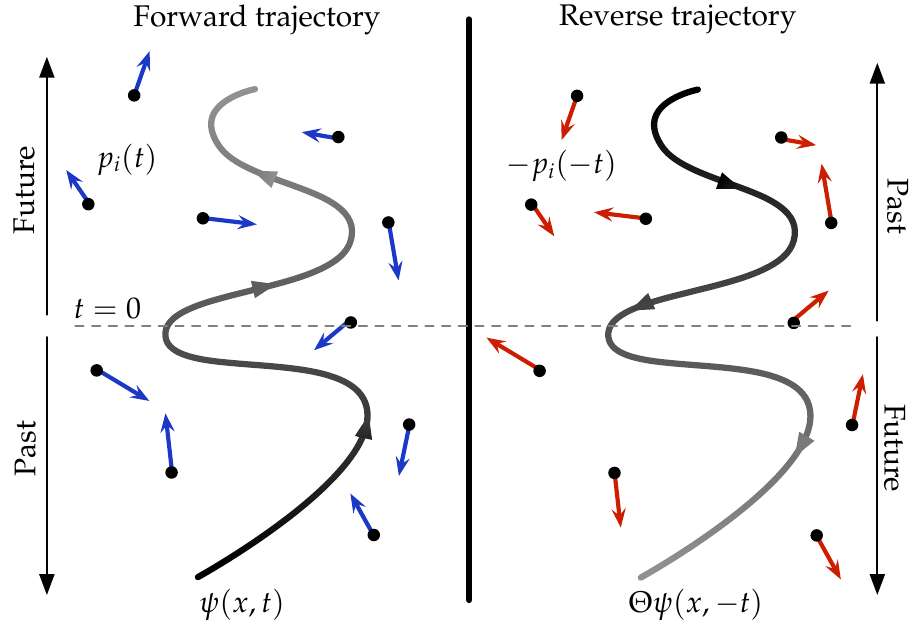}
\caption{Schematic showing the concept behind the time-reversal transformation. 
The system and environment move backward in time.
In quantum mechanics this is implemented using the time-reversal operator, which reverses the momenta, followed by the transformation $t \to -t$.} \label{TRFig}
\end{figure}

If we consider an equation of motion for an observable $\hat{A}(t)$ in the Heisenberg picture, then we say it is time-reversal symmetric if it is also satisfied by the operator $\hat{A}_{R}(t)$, which is the time-reversed counterpart to $\hat{A}$,
\begin{equation}
\hat{A}_{R}(t) = \hat{\Theta} \hat{A}(-t) \hat{\Theta}^{-1}.
\end{equation}
If a unitary evolution is governed by a time-independent Hamiltonian $\hat{H}$ which commutes with the time-reversal operator $\hat{\Theta} \hat{H}\hat{\Theta}^{-1} = \hat{H}$ (which is the case in \eqref{eq:H}), then in the Heisenberg picture we have
\begin{subequations}
\begin{align}
&\hat{\bm{q}}_{R}(t)=\hat{\Theta} \hat{\bm{q}}(-t) \hat{\Theta}^{-1} = \hat{\bm{q}}(t),\\
&\hat{\bm{p}}_{R}(t)=\hat{\Theta} \hat{\bm{p}}(-t) \hat{\Theta}^{-1} = -\hat{\bm{p}}(t). \quad 
\end{align}
\end{subequations}
As an example, we can show that the generalised Langevin equation \eqref{eq:GLE} is time-reversal symmetric. This is to be expected, as it is an exact evolution equation in the Heisenberg picture governed by a Hamiltonian which commutes with the time-reversal operator. 
To check that this is indeed the case \eqref{eq:GLE} implies the following equation of motion for $\hat{\bm{q}}(-t)$, 
\begin{align}
M\frac{d^{2}\hat{Q}}{dt^{2}}(-t) &+ V^{\prime}(\hat{Q}(-t))+\frac{1}{M}\int_{0}^{-t} k(-t-t^{\prime})\hat{P}(t^{\prime})\, dt^{\prime} \nonumber \\
&+ k(-t)\hat{Q}(0) = \hat{f}(-t).
\end{align}
Conjugating both sides of this equation with the time-reversal operator $\Theta (\cdot)\Theta^{-1}$ will return this equation back to \eqref{eq:GLE}, demonstrating its time-reversal symmetry.
Here we used the fact that the memory kernel \eqref{eq:kernel} is even $k(t) = k(-t)$.
Likewise the force term $f(t)$, despite having been reinterpreted as stochastic, was deterministic in origin, and here we treat it as such. Therefore, the time-reversed force term results from time-reversing the initial conditions of the bath $\hat{q}_{k}(0)$ and $\hat{p}_{k}(0)$, and hence it satisfies $\hat{\Theta}\hat{f}(t)\hat{\Theta}^{-1} = \hat{f}(-t)$.

For equations in the Schrödinger picture such as the Lindblad master equation \eqref{eq:Lindblad}, one needs to calculate the evolution equation for the time reversed density operator
$\hat{\rho}_{R}(t) = \hat{\Theta} \hat{\rho}(-t)\hat{\Theta}^{-1}$.
In particular for the Lindblad equation,
\begin{equation}
\frac{d\hat{\rho}_{R}}{dt} = -\hat{\Theta} \frac{d\hat{\rho}}{dt}\hat{\Theta}^{-1} = -\hat{\Theta}\hat{\mathcal{L}}\hat{\Theta}^{-1}\hat{\Theta}\hat{\rho}\hat{\Theta}^{-1} = -\hat{\mathcal{L}}_{R}\hat{\rho}_{R}.\label{eq:LindbladTR}
\end{equation}
So the Lindblad equation is time-reversal symmetric if the Lindbladian satisfies $\hat{\mathcal{L}} = -\hat{\mathcal{L}}_{R}$.
When the Lindblad equation reduces to the time-independent Schrödinger equation, this condition corresponds to the Hamiltonian commuting with the time-reversal operator.
The solution of \eqref{eq:LindbladTR} is described by a one-parameter family of CPTP maps $\set{\hat{\mathcal{E}}(t)}$:
\begin{equation}
\hat{\rho}(t) = \hat{\mathcal{E}}(t)\hat{\rho}(0), \qquad \hat{\mathcal{E}}(t) = \exp\left(-\hat{\mathcal{L}}t\right).
\end{equation}
By expanding $\hat{\Theta}\hat{\mathcal{E}}(t)\hat{\Theta}^{-1}$, we see that
\begin{equation}
\hat{\Theta}\hat{\mathcal{E}}(t)\hat{\Theta}^{-1} = \exp(-\hat{\Theta}\hat{\mathcal{L}}\hat{\Theta}^{-1}\, t) = \exp(-\hat{\mathcal{L}}_{R}t).
\end{equation}
If the system is time-reversal symmetric, then a necessary condition on the family of CPTP maps is
\begin{equation}
\hat{\Theta}\hat{\mathcal{E}}(t)\hat{\Theta}^{-1}= \exp(\hat{\mathcal{L}}t) = \hat{\mathcal{E}}(t)^{-1}.\label{eq:TRinvert}
\end{equation}
In the case in which the system evolves unitarily according to a time-independent Hamiltonian, then the family of CPTP maps reduces to a family of unitary maps $\set{\hat{U}(t)}$, and \eqref{eq:TRinvert} leads to the well-known condition for time-reversal symmetric unitary evolution \cite{Sakurai2020},
\begin{equation}
\hat{\Theta}\hat{U}(t) \hat{\Theta}^{-1} = \hat{U}(-t).
\end{equation}
More generally, if the generator $\hat{\mathcal{L}}$ is time-dependent, with an equation of motion of the form
\begin{equation}
\frac{d\hat{\rho}}{dt} = \hat{\mathcal{L}}(t)\hat{\rho}(t),\label{eq:genMR}
\end{equation}
then the time-reversed equation of motion is
\begin{equation}
\frac{d\hat{\rho}_{R}}{dt}(t) = -\hat{\Theta} \frac{d\hat{\rho}}{dt}(-t)\hat{\Theta}^{-1} = -\hat{\mathcal{L}}_{R}(-t)\hat{\rho}_{R}(t),
\end{equation}
In this case, the equation of motion is time-reversal symmetric if $\hat{\mathcal{L}}_{R}(t)=\hat{\Theta}\hat{\mathcal{L}}(t)\hat{\Theta}^{-1} = -\hat{\mathcal{L}}(-t)$.
By contrast with \eqref{eq:LindbladTR}, the solution of \eqref{eq:genMR} is described by invertible two-parameter family of CPTP maps $\set{\hat{\mathcal{E}}(t_{2},t_{1})}$:
\begin{equation}
\hat{\rho}(t) = \hat{\mathcal{E}}(t,0)\hat{\rho}(0), \quad \hat{\mathcal{E}}(t_{2},t_{1}) = \hat{\mathcal{T}}\exp\left(-\int_{t_{1}}^{t_{2}} \hat{\mathcal{L}}(t^{\prime})\, dt^{\prime}\right),
\end{equation}
where $\hat{\mathcal{T}}$ indicates the time-ordering operator.
Using the Dyson series expansion, we have
\begin{align}
\hat{\Theta}\hat{\mathcal{E}}(t_{1},t_{2})\hat{\Theta}^{-1} &= \hat{\mathcal{T}}\exp\left(-\int_{t_{1}}^{t_{2}}\hat{\Theta}\hat{\mathcal{L}}(t^{\prime})\hat{\Theta}^{-1}\, dt^{\prime}\right) \nonumber \\
&= \hat{\mathcal{T}}\exp\left(-\int_{t_{1}}^{t_{2}}\hat{\mathcal{L}}_{R}(t^{\prime})\, dt^{\prime}\right).
\end{align}
The requirement of time-reversal symmetry places the following necessary condition on the family of CPTP maps,
\begin{align}
\hat{\Theta}\hat{\mathcal{E}}(t_{1},t_{2})\hat{\Theta}^{-1} &= \mathcal{T}\exp\left(-\int_{t_{2}}^{t_{1}}\hat{\mathcal{L}}(t^{\prime})\, dt^{\prime}\right) \nonumber \\
&= \mathcal{E}(t_{2},t_{1})^{-1}.
\end{align}
So while the time-reversal symmetry is a condition on the equation of motion, in the case of Markovian evolution this symmetry imposes necessary conditions on the corresponding family of CPTP maps that describe the evolution of the system.

Demonstrating time-reversal symmetry for equations of motion can involve subtleties which require particular care.
Such equations frequently involve constants which are derived from tracing over environmental degrees of freedom.
This means that to calculate the time-reversed trajectory of the global system, one cannot simply treat these as constants, they must be time-reversed themselves.
In short, the time-reversal operator does not commute with the partial trace operation.
An example of this subtlety will occur in section \ref{sec:Lindblad}.\\

\subsection{The Markov Approximation and the symmetric quantum Langevin equation}

In order to derive a Markovian Langevin equation, we now introduce suitable approximations to deconvolve the convolution integral in the generalised Langevin equation \eqref{eq:GLE}.

This is usually achieved by assuming that the bath equilibrates within a time scale, which is much shorter than the time scale over which the system  significantly changes. Namely, we assume that
\begin{equation}
\int_{0}^{\infty} k(t^{\prime})\, dt^{\prime} < \infty, \label{intk}
\end{equation}
and that there exists some time scale $\tau_{\textsc{B}} \geq 0$ such that
\begin{equation}
\int_{0}^{\tau_{B}} k(t^{\prime})\, dt^{\prime} \approx \int_{0}^{\infty} k(t^{\prime})\, dt^{\prime}.
\end{equation}
We also assume that the momentum $\hat{P}(t)$ changes sufficiently slowly, such that on any temporal interval of length $\tau_{B}$ it is approximately constant.
Therefore, for $t \gtrsim \tau_{B}$,
\begin{align}
&\int_{0}^{t} k(t-t^{\prime}) \frac{dQ}{dt^{\prime}}(t^{\prime}) \, dt^{\prime} = \int_{0}^{t} k(t^{\prime}) \frac{dQ}{dt^{\prime}}(t-t^{\prime}) \, dt^{\prime}\nonumber \\
&\approx \frac{dQ}{dt}(t) \lim_{t\rightarrow \infty} \int_{0}^{t} k(t^{\prime})dt^{\prime} = \frac{dQ}{dt}(t) \int_{0}^{\infty} k(t^{\prime}) \, dt^{\prime},\label{eq:approx}
\end{align}

This standard implementation of the Markov approximation is only applicable for times $t>0$. Since the kernel $k(t)$ is an even function of time,
\begin{equation}
\int_{0}^{t} k(t^{\prime}) \, dt^{\prime} = \sgn(t) \int_{0}^{|t|} k(t^{\prime}) \, dt^{\prime},
\end{equation}
and therefore the correct approximation for $|t| \gtrsim \tau_{B}$ is
\begin{equation}
\int_{0}^{t} k(t-t^{\prime})\hat{P}(t^{\prime}) dt^{\prime} \approx \sgn(t)\hat{P}(t)  \int_{0}^{\infty} k(t^{\prime})dt^{\prime},\label{eq:revM}
\end{equation} 
a result that was remarked upon in \cite{Hanggi2005} in the case in which $k(t)\propto \delta(t)$.

Implementing this approximation to the generalised Langevin equation \eqref{eq:GLE} immediately leads to the following non-standard from of Langevin dyanmics, 
\begin{equation}
M \frac{d^{2}\hat{Q}}{dt^{2}} + V^{\prime}(\hat{Q}(t)) +  \sgn(t)\gamma \hat{P}(t) = \hat{f}(t).\label{eq:qsgnlang}
\end{equation}
where we have introduced the dissipation constant $\gamma$ by re-writing,
\begin{equation}
\int_{0}^{\infty} k(t^{\prime})\, dt^{\prime} = M \gamma.
\end{equation}
Equation \eqref{eq:qsgnlang} can be checked to be time-reversal symmetric using the time-reversal operator, and it represents the central observation of this paper: that properly applying the Markov condition into the past produces a time-reversal symmetric equation of motion, where, in contrast to the standard quantum Langevin equation, the function $\sgn(t)$ in Eq. \eqref{eq:qsgnlang} cancels the sign change incurred by the velocity term under the time-reversal transformation. For this reason, thermalisation is predicted not only into the future but also into the past. 

It is worth emphasising that the preservation of the time-reversal symmetry is due to the memory kernel being an even function of time. The Markov approximation is typically implemented on physical grounds by carefully choosing the distribution of bath frequencies and coupling constants to arrive at an integrable memory kernel, as in Eq. (\ref{intk}).
Most common is the Ohmic approximation, in which the bath is approximated by a continuum of frequencies with a linear spectral density so that the memory kernel is a Dirac delta function \cite{BreuerPetruccione07}. Since this is still an even function, \eqref{eq:qsgnlang} still follows.
We note that in the Ohmic approximation the quantum correlation function \eqref{eq:qcorr} becomes, 
\begin{align}
&\expect{\set{\hat{f}(t),\hat{f}(t^{\prime})}} =\nonumber \\
&\qquad\frac{2\gamma M\hbar}{\pi}\int_{0}^{\Lambda} \omega \coth\left(\frac{\hbar \omega}{2k_{B}T}\right)\cos(\omega(t-t^{\prime})) d\omega, \label{eq:qcorrohm}
\end{align}
where $\Lambda$ is the largest bath oscillator frequency.
In the high-temperature limit $k_{B} T \gg \hbar \omega$, and large cutoff limit, the autocorrelation function is Dirac-delta correlated,
\begin{equation}
\expect{\set{\hat{f}(t),\hat{f}(t^{\prime}}} = 2\gamma M k_{B} T \delta(t-t^{\prime}). \label{eq:qcorrhighT}
\end{equation} 

The evenness of $k(t)$ will also hold for all non-Ohmic choices of the spectral density, and therefore there is no such choice that could lead to an equation of motion that violates time-reversal symmetry.

\subsection{The time-reversal symmetric quantum master equation}\label{sec:CL}

The quantum Brownian motion master equation can be derived from \eqref{eq:qsgnlang} by tracing over the bath Hilbert space. Here we will continue to follow the approach presented in \cite{Gardiner2004}, highlighting in particular the deviations from the standard case due to the time-symmetric nature of the quantum Langevin equation (\ref{eq:qsgnlang}).

The first step is to partially transform back into the Schrödinger picture by defining the adjoint operator $\hat{\mu}(t)$, acting on both system and environment Hilbert spaces, implicitly through the equation 
\begin{equation}
\ptr{\textsc{S}}{\hat{Y}\hat{\mu}(t)} = \ptr{\textsc{S}}{\hat{\rho}_\textsc{S}\hat{Y}(t)},
\end{equation}
where $\hat{Y}$ is any Schrödinger picture system operator, and $\hat{Y}(t)$ is its Heisenberg picture counterpart. Here we have assumed that in the Heisenberg picture the state of the system and environment factorises into $\hat{\rho}_{\textsc{S}}\otimes \hat{\rho}_{\textsc{B}}$. From Heiseinberg's equation, $\hat{\mu}(t)$ can be shown to satisfy the following differential equation,
\begin{align}
\dot{\hat{\mu}}(t) = -\frac{i}{\hbar} \left[\hat{H}_{\textsc{S}},\hat{\mu}(t)\right] &+ \frac{i}{2\hbar} \left[\set{\gamma\, \sgn(t) \hat{P},\hat{\mu}(t)},\hat{Q}\right]\nonumber \\
&+ \frac{i}{2\hbar} \left[\set{\hat{f}(t),\hat{\mu}(t)},\hat{Q}\right].\label{eq:adjoint}
\end{align}
where the $\sgn(t)$ factor carries through the derivation from the quantum Langevin equation. We note that this differential equation is time-reversal symmetric.

To derive the corresponding quantum master equation, we define the density operator as 
\begin{equation}
\hat{\rho}(t) = \expect{\hat{\mu}(t)} := \ptr{\textsc{B}}{\hat{\mu}(t)\hat{\rho}_{\text{th}}},
\end{equation}
where $\hat{\rho}_{\text{th}} = \hat{\rho}_{\textsc{B}}$ is the initial state of the bath, thermal with respect to the Hamiltonian $\hat{H}_{\textsc{B}}$.

The trace over the bath can be done following van Kampen's cumulant expansion \cite{vanKampen2007,Gardiner2004} (see appendix \ref{app:cumulant}), and results in the master equation
\begin{align}
\dot{\hat{\rho}}(t) &= -\frac{i}{\hbar} \left[\hat{H}_{\textsc{S}},\hat{\rho}(t)\right] + \frac{i\,\sgn(t)}{2\hbar} \left[\set{\gamma\,  \hat{P},\hat{\rho}(t)},\hat{Q}\right] \nonumber \\
&\qquad- \frac{\Gamma(t)}{\hbar^{2}} \left[\left[\hat{\rho}(t),\hat{Q}\right],\hat{Q}\right].\label{eq:masterapprox}
\end{align}
The coefficient $\Gamma(t)$ is given by the integral of the autocorrelation function \eqref{eq:qcorrohm}
\begin{equation}
\Gamma(t) = \int_{0}^{t} \expect{\set{\hat{f}(t),\hat{f}(t^{\prime})}}dt^{\prime} \label{eq:gamma}
\end{equation}
At this stage of the derivation of the master equation \eqref{eq:masterapprox}, the time-reversal symmetry has not been violated. The violation of time-reversal symmetry typically occurs at this point by implementing the Markov approximation: that is approximating $\Gamma(t) \approx \Gamma(\infty)$ for all $t$ greater than the correlation time of the bath. However this artificially introduces an arrow of time, as  $\expect{\set{\hat{f}(t),\hat{f}(t^{\prime})}}$ given by \eqref{eq:qcorr} is an even function of $t-t^{\prime}$, and therefore 
\begin{equation}
\int_{0}^{t} \expect{\set{\hat{f}(t),\hat{f}(t^{\prime})}} dt^{\prime} = \sgn(t)\int_{0}^{|t|} \expect{\set{\hat{f}(|t|),\hat{f}(t^{\prime})}} dt^{\prime}
\end{equation}
must hold. 
In the limit $\Lambda\rightarrow\infty$, together with the assumption of high temperature, $k_B T \gg \hbar \omega$ \eqref{eq:qcorrhighT}, we obtain
\begin{equation}
\lim_{t\rightarrow \infty} \int_{0}^{|t|} \expect{\set{\hat{f}(|t|),\hat{f}(t^{\prime})}}dt^{\prime} = 2\gamma M k_{B} T.
\end{equation}
Therefore, the appropriate approximation, for both positive and negative times, leads to the master equation
\begin{align}
\dot{\hat{\rho}}(t) &= -\frac{i}{\hbar} \left[\hat{H}_{\textsc{S}},\hat{\rho}(t)\right] + \frac{i\,\sgn(t)}{2\hbar} \left[\set{\gamma\,  \hat{P},\hat{\rho}(t)},\hat{Q}\right] \nonumber \\
&\quad- \sgn(t)\frac{2\gamma M k_{B} T}{\hbar^{2}} \left[\left[\hat{\rho}(t),\hat{Q}\right],\hat{Q}\right]. \label{eq:BMsol}
\end{align}
Equation (\ref{eq:BMsol}) is time-reversal symmetric, as can be checked using the time-reversal operator. 
We recast it into the form
\begin{equation}
\frac{d\hat{\rho}}{dt}(t) = (i\hat{\mathcal{L}}_{H} + \sgn(t)\hat{\mathcal{L}}_{D})\hat{\rho}(t), \label{msl}
\end{equation}
with 
\begin{subequations}
\begin{eqnarray}
&&\hat{\mathcal{L}}_{H} \hat{\rho}(t) = -\frac{1}{\hbar} \left[\hat{H}_{\textsc{S}},\hat{\rho}(t)\right], \\
&&\hat{\mathcal{L}}_{D}\hat{\rho}(t) = \frac{i}{2\hbar} \left[\set{\gamma\,  \hat{P},\hat{\rho}(t)},\hat{Q}\right] \nonumber \\
&&\qquad \qquad \qquad - \frac{2 \gamma M k_{B} T}{\hbar^{2}} \left[\left[\hat{\rho}(t),\hat{Q}\right],\hat{Q}\right],
\end{eqnarray}
\end{subequations}
It is simple to check that $\hat{\Theta}\hat{\mathcal{L}}_{H}\hat{\Theta}^{-1} = \hat{\mathcal{L}}_{H}$ and similarly $\hat{\Theta}\hat{\mathcal{L}}_{D}\hat{\Theta}^{-1} = \hat{\mathcal{L}}_{D}$, and therefore
\begin{equation}
\frac{d\hat{\rho}_{R}}{dt}(t) = -\hat{\Theta}\frac{d\hat{\rho}}{dt}(t)\hat{\Theta}^{-1} = (i\hat{\mathcal{L}}_{H} + \sgn(t)\hat{\mathcal{L}}_{D})\hat{\rho}_{R}(t),
\end{equation}
The solution, can be written in the form
\begin{equation}
\hat{\rho}(t) = \hat{\mathcal{E}}(t)
\hat{\rho}(0), \qquad \hat{\mathcal{E}}(t) = \exp\left(i \hat{\mathcal{L}}_{H}t + \hat{\mathcal{L}}_{D}|t|\right),
\end{equation}
which shows that it extends the standard quantum Brownian motion master equation for negative times.
Without the factor of $\sgn(t)$, \eqref{eq:BMsol} is simply the standard Brownian motion master equation, which violates time-reversal symmetry.
The solution is described by an exponential map which clearly satisfies the divisibility criterion \eqref{eq:divisible} into the future $t_{1},t_{2} > 0$ and into the past $t_{1},t_{2}<0$, with respect to the time origin $t=0$.

While this master equation has widely been seen as the quantum analogue of classical Brownian motion, it can be shown that this master equation corresponds to an evolution map which is not completely-positive \cite{Homa2019}, and it has been argued that the correct Markovian master equation for this system should not even be dissipative \cite{Ferialdi2017}.
As such it does not meet the technical definition of Markovian evolution described in section \ref{sec:markov}.
Nevertheless it is a very frequently used master equation which is derived from an open system model by invoking several Markovian approximations.
Our analysis shows that these approximations do not destroy the time-reversal symmetry that exists in the full model.

\subsection{Entropy}

We illustrate the effect of the $\sgn(t)$ factors in Eq. (\ref{eq:BMsol}) by plotting the behaviour of the von Neumann entropy over time.

The solution of \eqref{eq:BMsol} can be computed by simply adapting the solution found in \cite{Venugopalan1994}. Explicitly, in the momentum representation, this can be written as 
\begin{equation}
\rho(p,q,t) = \frac{1}{\sqrt{\pi N(t)}}\exp\left(\frac{-(q+\sgn(t)A(t)p)^{2}}{N(t)} - B(t)p^{2}\right),
\end{equation}
where here $p = u-v$ and $q=(u+v)/2$ are the rotated momentum coordinates, and
\begin{subequations}
\begin{align}
N(t) &= \frac{m k_{B} T}{\hbar^{2}}\left(1-e^{-2\gamma |t|}\right) + \frac{e^{-2\gamma |t|}}{\sigma^{2}}, \\
A(t) &= \frac{i\hbar}{2\sigma^{2}m \gamma}e^{-\gamma |t|}(1-e^{-\gamma|t|}) -\frac{ik_{B}T}{2\hbar \gamma}(1-e^{-\gamma|t|})^{2}, \\
B(t) &= \frac{\hbar^{2}}{4\sigma^{2}m^{2}\gamma^{2}}(1-e^{-\gamma |t|)})^{2} + \frac{\sigma^{2}}{4} \nonumber \\
& \quad +\frac{k_{B}T}{m\gamma^{2}}(2\gamma |t|-3+4e^{-\gamma |t|}-e^{-2\gamma |t|}).
\end{align}
\end{subequations}
\\
At $t=0$ the state is a pure Gaussian wavefunction with variance $\sigma^{2}/2$,
\begin{equation}
\psi(x,0) = \frac{1}{(\sigma^{2}\pi)^{1/4}} \exp\left(-\frac{x^{2}}{2\sigma^{2}}\right). 
\end{equation}
The von Neumann entropy of a Gaussian state can be calculated from the purity $\xi = \tr(\rho^{2})$ using the formula \cite{Horhammer2008}
\begin{equation}
S_{\text{vN}}(\xi) = \frac{1-\xi}{2\xi}\log\left(\frac{1+\xi}{1-\xi}\right) - \log\left(\frac{2\xi}{1+\xi}\right).
\end{equation}

\begin{figure}[h!]
\centering
\includegraphics[scale=0.45]{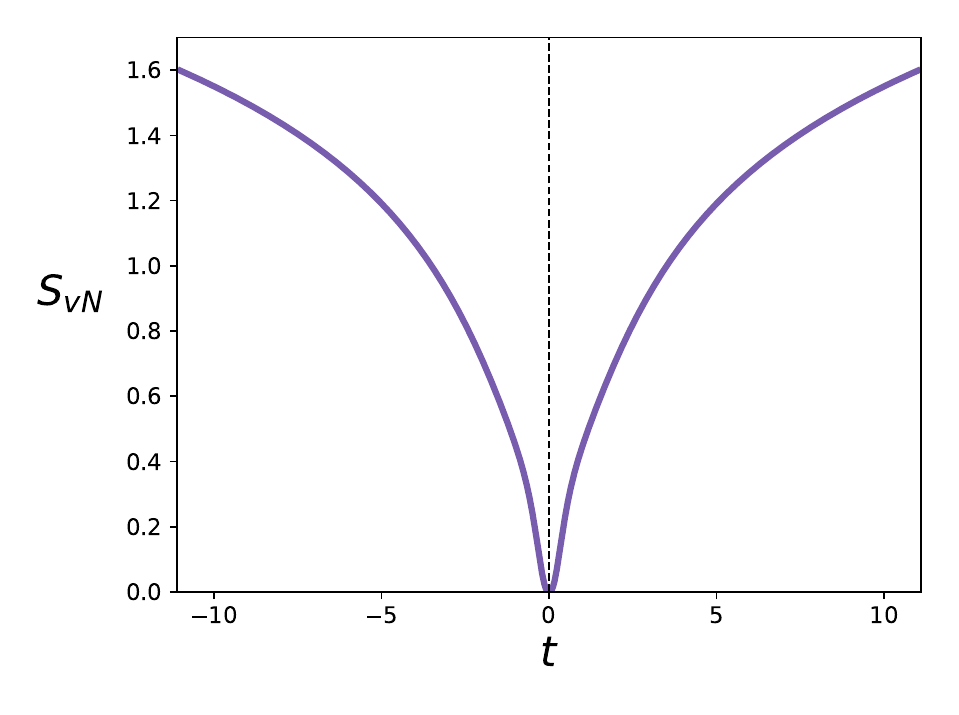}
\caption{The von Neunmann entropy $S_{\textsc{vN}}$ of the solution \cite{Venugopalan1994} to the time-symmetric quantum Brownian motion master equation \eqref{eq:BMsol}, in a pure state at $t=0$. Here $\gamma=M=k_{B}T = \hbar = 1$. The von Neumann entropy is calculated from the purity \cite{Horhammer2008}, and increases unbounded as the off-diagonal terms of the density operator in the momentum representation decay to zero. The entropy increases (irreversibility) in both temporal directions but there is nonetheless no arrow-of-time, as time-reversal symmetry is maintained. Increasing entropy only indicates motion away from the temporal origin, not whether system is moving forwards or backwards in time.}\label{fig:qEnt}
\end{figure}

The plot in Fig.~\ref{fig:qEnt} shows the von Neumann entropy corresponding to the solution of \eqref{eq:BMsol}. The time evolution is symmetric with respect to the origin of time, as a result of the $\sgn(t)$ factors in \eqref{eq:BMsol}. Once a direction of time has been fixed {\it a priori}, these dynamics describe standard irreversibility, associated to the standard increase of entropy. However, from the origin of time, an opposite arrow of time emerges, which describes irreversibility towards the past as well.
This plot demonstrates the clear distinction between the concepts of irreversibility and time-reversal symmetry as described in section~\ref{thetasec}.
In both temporal directions the system is evolving non-unitarily, irreversibly losing information to the environment as entropy increases.
Nevertheless, there is no arrow-of-time in this system, since the increase of entropy does not distinguish between the forward or backward trajectory. The increase of entropy only indicates movement away from the temporal origin.

\section{Symmetric Lindblad Master Equation\label{sec:Lindblad}}

The Lindblad equation is the most general form of quantum master equation for the Markovian evolution of a quantum system \cite{Gorini76, Lindblad76}, and it is typically not time-reversal symmetric. In fact, it can be seen from \eqref{eq:Lindblad} that it is not when the Lindbladian $\hat{\mathcal{L}}$ commutes with the time-reversal operator.
In this section we examine the role of the Markov approximation in the standard microscopic derivation of the Lindblad master equation \cite{BreuerPetruccione07}.

Consider again a system and environment described by the total time-independent Hamiltonian
\begin{equation}
\hat{H} = \hat{H}_{\textsc{S}} + \hat{H}_{\textsc{B}} + \hat{H}_{\rm{SB}}.\label{eq:fullH}
\end{equation}
While in this case we do not make any assumption on the explicit form of the Hamiltonians $\hat{H}_{\textsc{S}}$, $\hat{H}_{\textsc{B}}$ and $\hat{H}_{\rm{SB}}$, we assume that all three Hamiltonian terms commute with the time-reversal operator.

In the interaction picture the dynamics is governed by the time-dependent Hamiltonian
\begin{equation}
\hat{H}_{\textsc{I}}(t) = e^{\frac{i}{\hbar}(\hat{H}_{\textsc{S}}+\hat{H}_{\textsc{B}})t}\hat{H}_{\textsc{SB}}\,e^{\frac{-i}{\hbar}(\hat{H}_{\textsc{S}}+\hat{H}_{\textsc{B}})t}.\label{eq:intH}
\end{equation}
Since the time-reversal operator commutes with all three Hamiltonian terms in (\ref{eq:fullH}), then the von Neumann equation in this picture is time-reversal symmetric,
\begin{equation}
\frac{d}{dt}\hat{\rho}_{\textsc{I}}(t) = -\frac{i}{\hbar}\left[\hat{H}_{\textsc{I}}(t),\hat{\rho}_{\textsc{I}}(t)\right], \label{vn}
\end{equation}
where $\hat{\rho}_{\textsc{I}}(t)$ is the density operator in the interaction picture.

By formally solving the von Neumann equation (\ref{vn}), substituting that solution into itself, and tracing over the environment with the assumption $\ptr{\textsc{B}}{[\hat{H}_{\textsc{I}}(t),\hat{\rho}_{\textsc{I}}(0)]}=0$,
%\begin{equation}
%\ptr{\textsc{B}}{[H_{\textsc{I}}(t),\rho_{\textsc{I}}(0)]}=0,
%\end{equation}
we obtain
\begin{equation}
\frac{d}{dt}\hat{\rho}_{\textsc{S}}(t) =  - \frac{1}{\hbar^{2}}\int_{0}^{t} \ptr{\textsc{B}}{[\hat{H}_{\textsc{I}}(t),[\hat{H}_{\textsc{I}}(s),\hat{\rho}_{\textsc{I}}(s)]]} \,ds,\label{eq:dblint}
\end{equation}
where we have defined $\hat{\rho}_{\textsc{S}}(t) = \ptr{\textsc{B}}{\hat{\rho}_{\textsc{I}}(t)}$.

It is at this point that the Born-Markov approximation is made  \cite{BreuerPetruccione07} as follows. We first assume that the system and the bath are always approximately in a product state $\hat{\rho}_{\textsc{I}}(t) \approx \hat{\rho}_{\textsc{S}}(t)\otimes \hat{\rho}_{\textsc{B}}$. 
Moreover, we also assume that there is a natural timescale $\tau_{\textsc{B}}>0$ such that the integrand of \eqref{eq:dblint} effectively vanishes for all $|t|>\tau_{B}$, so that on this timescale $\hat{\rho}_{\textsc{S}}(t)$ is approximately constant, and therefore we have for $|t|>\tau_{\textsc{B}}$ 
\begin{equation}
\frac{d}{dt}\hat{\rho}_{\textsc{S}}(t) =  - \frac{1}{\hbar^{2}}\int_{0}^{t} \ptr{\textsc{B}}{[\hat{H}_{\textsc{I}}(t),[\hat{H}_{\textsc{I}}(s),\hat{\rho}_{\textsc{S}}(t)\otimes \hat{\rho}_{\textsc{B}}]]} \,ds. \label{eq:micro}
\end{equation}
Equation (\ref{eq:micro}) is the so-called Redfield equation \cite{Redfield57}, which describes a dynamics which is local in time, even though not Markovian, as the upper bound of the integral in the RHS is still $t$ \cite{BreuerPetruccione07}. It is worth emphasizing that because of this dependence, the Redfield equation is still time-reversal symmetric.

It is then standard to approximate the upper limit of the integral as $\infty$ and to bring the equation into Lindblad form. However, this approximation produces an artificial violation of the time-reversal symmetry.
To analyse this more carefully we will re-write the Redfield equation into Lindblad form before applying the Markov approximation.

We proceed by making the secular approximation by re-writing $\hat{H}_{\textsc{I}}(t)$ in terms of local operators. Assuming a discrete spectrum for the local system Hamiltonian $\hat{H}_{\textsc{S}}$, 
we can write
\begin{equation}
\hat{H}_{\textsc{I}}(t) = \sum_{\alpha,\omega} e^{-i\omega t} \hat{A}_{\alpha}(\omega) \otimes \hat{B}_{\alpha}(t),
\end{equation}
where $\hat{A}_{\alpha}(\omega)e^{-i\omega t}$ is the system interaction component $\hat{A}_{\alpha}(t)$ projected onto the eigenstates of $H_{\textsc{S}}$ with frequency difference $\omega$. With the trace over the environment, the master equation \eqref{eq:micro} can now be written
\begin{align}
\frac{d}{dt}&\hat{\rho}_{\textsc{S}}(t) =  -\frac{1}{\hbar^{2}}\sum_{\alpha,\beta,\omega}\left[\Gamma_{\alpha\beta}(\omega,t)\hat{A}_{\alpha}^{\dagger}(\omega)\hat{A}_{\beta}(\omega)\hat{\rho}_{\textsc{S}}(t)\right. \nonumber \\
&\qquad \qquad \qquad \quad + \Gamma^{*}_{\beta\alpha}(\omega,t) \hat{\rho}_{\textsc{S}}(t)\hat{A}_{\alpha}^{\dagger}(\omega)\hat{A}_{\beta}(\omega)\nonumber \\
&- \left.\left(\Gamma_{\alpha\beta}(\omega,t) + \Gamma^{*}_{\beta\alpha}(\omega,t)\right)\hat{A}_{\beta}(\omega)\hat{\rho}_{\textsc{S}}(t)\hat{A}_{\alpha}^{\dagger}(\omega)\right].\label{eq:Lbd}   
\end{align}
The secular approximation was done here by replacing all terms of the form $\exp(i(\omega^{\prime}-\omega)t)$ with the Kronecker delta $\delta_{\omega,\omega^{\prime}}$.
The coefficients $\Gamma_{\alpha\beta}$ are defined as
\begin{equation}
\Gamma_{\alpha\beta}(\omega,t) = \int_{0}^{t	} e^{i\omega s}\, \tr\left(\hat{B}^{\dagger}_{\alpha}(t)\hat{B}_{\beta}(t-s)\hat{\rho}_{\textsc{B}}\right)\, ds.
\end{equation}

This master equation is still time-reversal symmetric after the secular approximation.
If $\hat{\rho}_{\textsc{B}}$ is a stationary state $[\hat{H}_{\textsc{B}},\hat{\rho}_{\textsc{B}}]=0$, such as a thermal state, then the autocorrelation function is stationary. As a consequence, we have
\begin{align}
\Gamma_{\alpha\beta}(\omega,t) &+ \Gamma^{*}_{\beta\alpha}(\omega,t) = \\
&\sgn(t)\int_{-|t|}^{|t|} e^{i\omega s} \, \tr\left(\hat{B}^{\dagger}_{\alpha}(s)\hat{B}_{\beta}(0)\hat{\rho}_{B}\right)\, ds.\nonumber
\end{align}

We now approximate the limits of the integral as infinite, for times $|t| >\tau_{B}$. We thus now write
\begin{equation}
  \Gamma_{\alpha\beta}(\omega,t) + \Gamma^{*}_{\beta\alpha}(\omega,t) \approx \sgn(t)\gamma_{\alpha\beta}(\omega)
\end{equation}
where 
\begin{equation}
\gamma_{\alpha\beta}(\omega) = \int_{-\infty}^{\infty} e^{i\omega s} \, \tr\left(\hat{B}^{\dagger}_{\alpha}(s)\hat{B}_{\beta}(0)\hat{\rho}_{B}\right)\, ds.
\end{equation}

Similarly, the difference $\Gamma_{\alpha\beta}(\omega,t) - \Gamma^{*}_{\beta\alpha}(\omega,t)$ is an even function of $t$, so it becomes time-independent when the integration limits are taken to infinity, and therefore we can use the standard definition 
\begin{equation}
\eta_{\alpha\beta}(\omega) = \frac{1}{2i}\left(\Gamma_{\alpha\beta}(\omega,t) - \Gamma^{*}_{\beta\alpha}(\omega,t)\right).
\end{equation}

With these definitions, the Lindblad master equation \eqref{eq:Lbd} can be rewritten in the non-conventional form
\begin{equation}
\frac{d}{dt}\hat{\rho}_{\textsc{S}}(t) = -\frac{i}{\hbar}[\hat{H}_{\textsc{S}},\hat{\rho}_{\textsc{S}}] + \frac{\sgn(t)}{\hbar^{2}}\sum_{\alpha,\beta,\omega} \mathcal{\hat{D}}_{\alpha\beta}(\omega)\hat{\rho}_{\textsc{S}}(t),\label{eq:Lbld}
\end{equation}
where we have defined the Hermitian operator
\begin{equation}
\hat{H}_{\textsc{S}} = \frac{1}{\hbar} \sum_{\alpha,\beta,\omega} \eta_{\alpha\beta}(\omega)\hat{A}_{\alpha}^{\dagger}(\omega)\hat{A}_{\beta}(\omega),
\end{equation}
as well as the dissipation superoperators
\begin{align}
\mathcal{\hat{D}}_{\alpha\beta}(\omega)\hat{\rho}_{\textsc{S}}(t) =\, &\gamma_{\alpha\beta}(\omega)\bigg( \hat{A}_{\beta}(\omega)\hat{\rho}_{\textsc{S}}(t)\hat{A}_{\alpha}^{\dagger}(\omega) \\
&-\frac{1}{2}\set{\hat{A}^{\dagger}_{\alpha}(\omega)\hat{A}_{\beta}(\omega),\hat{\rho}_{\textsc{S}}(t)}\bigg).\nonumber
\end{align}
In order to analyze the time-reversal properties of Eq. \eqref{eq:Lbld} we note that the constants $\gamma_{\alpha,\beta}(\omega)$ and $\eta_{\alpha\beta}(\omega)$ must be treated carefully when performing the time-reversal transformation. 
They are defined in terms of traces over bath operators, which we must also reverse; just as one must reverse the initial conditions of the bath in the generalised Langevin equation. 
Hence:
\begin{align}
\hat{\Theta} \gamma_{\alpha\beta}(\omega)\hat{\Theta}^{-1} &= \int_{-\infty}^{\infty} e^{i\omega s} \, \tr\left(\hat{\Theta} \hat{B}^{\dagger}_{\alpha}(s)\hat{B}_{\beta}(0)\hat{\Theta}^{-1}\hat{\rho}_{B}\right)\, ds \nonumber \\
&= \gamma_{\alpha\beta}(\omega).
\end{align}
and similarly $\hat{\Theta} \eta_{\alpha\beta}(\omega)\hat{\Theta}^{-1} = \eta_{\alpha\beta}(\omega)$. Since $\hat{H}_{\textsc{I}}(t)$ commutes with the time-reversal operator, the superoperators $\mathcal{\hat{D}}_{\alpha\beta}(\omega)$ do also, which implies that \eqref{eq:Lbld} is time-reversal symmetric.

We therefore conclude that the form \eqref{eq:Lbld} of the Lindblad master equation  is time-reversal symmetric, and satisfies the symmetric Markov property. Any state which is a steady state of the future is also a steady state of the past.

\section{Symmetric Pauli Master Equation\label{sec:Pauli}}

The Pauli master equation describes the evolution of the diagonal terms of the density operator for a Markovian system. It is traditionally written as 
\begin{equation}
\frac{d}{dt} \rho_{n}(t) = \sum_{n^{\prime}\ne n} \left(W_{n,n^{\prime}}\rho_{n^{\prime}}(t) - W_{n^{\prime},n}\rho_{n}(t)\right), \label{eq:pauli}
\end{equation}
where $\rho_{n}(t) = \bra{n}{\hat{\rho}(t)}\ket{n}$ is the density operator population corresponding to the energy eigenstate $\ket{n}$, and where the transition rates $W_{n,n^{\prime}}$ are typically provided by Fermi's golden rule. This equation typically models the diagonal terms of a density matrix thermalising towards a steady state, and as such it is not time-reversal symmetric, as the transition rates are real numbers which commute with the time-reversal operator.

Equation (\ref{eq:pauli}) can be directly derived from the Lindblad master equation as introduced above, Eq. \eqref{eq:Lbld}, under the assumption that the system Hamiltonian $H_{\textsc{S}}$ is non-degenerate, namely $\hat{H}_{\textsc{S}} = \sum_{n} \varepsilon_{n} \ketbra{n}{n}$.
This, along with the fact that the local interaction operators $\hat{A}_{\alpha}(\omega)$ have been projected into the eigenstates of $\hat{H}_{\textsc{S}}$ implies that the off-diagonal terms of $\hat{A}_{\alpha}^{\dagger}(\omega)\hat{A}_{\beta}(\omega)$ vanish,
\begin{equation}
\bra{n}\hat{A}_{\alpha}^{\dagger}(\omega)\hat{A}_{\beta}(\omega)\ket{n^{\prime}} =  \delta_{n,n^{\prime}} \bra{n}\hat{A}_{\alpha}\ketbra{m}{m}\hat{A}_{\beta}\ket{n},
\end{equation}
where $\varepsilon_{m} = \varepsilon_{n} - \omega$.
As a consequence, when calculating the derivative of $\rho_{n}(t)$ from \eqref{eq:Lbld} the contribution from the Hamiltonian term is zero, and the contribution from the dissipation terms can be written in the form of \eqref{eq:pauli}.
However since our Lindblad master equation is time-reversal symmetric, the Pauli master equation is also time-reversal symmetric:
\begin{equation}
\frac{d}{dt} \rho_{n}(t) = \frac{\sgn(t)}{\hbar^{2}}\sum_{n\ne n^{\prime}} \left(W_{n,n^{\prime}}\rho_{n^{\prime}}(t) - W_{n^{\prime},n}\rho_{n}(t)\right),\label{eq:paulisym}
\end{equation}
where
\begin{equation}
W_{n^{\prime},n} = \sum_{\alpha,\beta} \gamma_{\alpha\beta}(\varepsilon_{n}-\varepsilon_{n^{\prime}})\bra{n}\hat{A}_{\alpha}\ketbra{n^{\prime}}{n^{\prime}}\hat{A}_{\beta}\ket{n}.
\end{equation}

It is also possible to derive a similar Pauli master equation directly from an open systems model by employing the Markov approximation  \cite{Pottier2010}. 
We begin with the Hamiltonian,
\begin{equation}
\hat{H} = \hat{H}_{0} + \lambda \hat{H}_{\textsc{SB}}.
\end{equation}
where $\lambda$ is the interaction strength and $\hat{H}_{0}$ is the local Hamiltonian for both the system and the bath, namely $\hat{H}_{0}=\hat{H}_{\rm S}+\hat{H}_{\rm B}$. For simplicity we will derive the master equation for the joint system and environment. The master equation for the system alone can be acquired by tracing over the environment.

Again we assume that both Hamiltonian terms commute with the time-reversal operator. We will index the eigenstates of $\hat{H}_{0}$ by $\ket{\varepsilon,k}$, where $\varepsilon$ is the energy value and $k$ tracks the degeneracy.
If $\lambda$ is sufficiently weak, then we can expand von Neumann equation up to second order in $\lambda$, to get the following differential equation for the diagonal terms \cite{Pottier2010},
\begin{align}
\frac{d}{dt}&\rho_{\varepsilon,k} (t) = \nonumber \\
&\frac{\lambda^{2}}{\hbar^{2}}\sum_{\varepsilon^{\prime},k^{\prime}}\int_{0}^{t} \Lambda_{\varepsilon,k,\varepsilon^{\prime},k^{\prime}}(t-s) (\rho_{\varepsilon^{\prime},k^{\prime}}(s) - \rho_{\varepsilon,k}(s))\, ds,
\end{align}
where $\rho_{\varepsilon,k}(s) = \bra{\varepsilon,k}\hat{\rho}(s)\ket{\varepsilon,k}$ and
\begin{equation}
\Lambda_{\varepsilon,k,\varepsilon^{\prime},k^{\prime}}(t) = 2\left|\bra{\varepsilon,k}\hat{H}_{\textsc{SB}}\ket{\varepsilon^{\prime},k^{\prime}}\right|^{2}\cos\left(\frac{\varepsilon-\varepsilon^{\prime}}{\hbar}t\right).
\end{equation}
It is assumed that $\rho(0)$ was diagonal in the $\ket{\epsilon,k}$ basis.

To Markovianise the system we now presume that the density operator populations evolve slowly (i.e. they are almost constant) over the time scale for which the integral is non-trivial.
This allows the integral to be evaluated over the memory kernel,
\begin{align}
\int_{0}^{t} \Lambda_{\varepsilon,k,\varepsilon^{\prime},k^{\prime}}&(t-s) \, ds = \nonumber \\
&2\hbar\frac{\left|\bra{\varepsilon,k}\hat{H}_{\textsc{SB}}\ket{\varepsilon^{\prime},k^{\prime}}\right|^{2}}{\varepsilon-\varepsilon^{\prime}}\sin\left(\frac{\varepsilon-\varepsilon^{\prime}}{\hbar}t\right).
\end{align}
Furthermore, we approximate the bath as being so large that the energy spectrum can be approximated by a continuum, with density $\eta(\varepsilon)$.
The master equation is still time-reversal symmetric at this stage.

Finally, for times $|t|$ larger than some characteristic time $\tau_{B}$ of the energy distribution, we can approximate
\begin{equation}
\frac{\hbar}{\varepsilon-\varepsilon^{\prime}}\sin\left(\frac{\varepsilon-\varepsilon^{\prime}}{\hbar}t\right) \approx \sgn(t)\pi \hbar\delta(\varepsilon-\varepsilon^{\prime}).
\end{equation}
The factor $\sgn(t)$ appears since $\sin(t)$ is an odd function of time.
Together, these approximations lead to the following form of Pauli master equation
\begin{equation}
\frac{d}{dt} \rho_{\varepsilon,k}(t) = \sgn(t)\sum_{k^{\prime}} \left(W^{\varepsilon}_{k,k^{\prime}}\rho_{\epsilon,k^{\prime}}(t) - W^{\epsilon}_{k^{\prime},k}\rho_{\epsilon,k}(t)\right), \label{paulieq}
\end{equation}
where $W^{\varepsilon}_{k,k^{\prime}} = W^{\varepsilon}_{k^{\prime},k}$ is given by
\begin{equation}
W^{\varepsilon}_{k,k^{\prime}} = \frac{2\lambda^{2}}{\hbar}\left| \bra{\varepsilon,k}\hat{H}_{\textsc{SB}} \ket{\varepsilon,k^{\prime}} \right|^{2} \eta(\varepsilon).
\end{equation}
Equation (\ref{paulieq}) is time-reversal symmetric.

\section{Breakdown of Time-Translation Symmetry}\label{timetransl}

The time-reversal symmetric quantum Langevin equation \eqref{eq:qsgnlang} exhibits exponential decay and relaxation into the past and into the future of the origin of time $t=0$. Thus, the origin becomes a particular point in which the past changes into the future.

As discussed in section~\ref{sec:markov}, Markovian dynamics forwards in time cannot be extrapolated indefinitely into the past as this would result in ill-defined density operators. In the open systems models analysed here, the origin thus represents this temporal boundary; from the point of view of the future evolution, it is the time at which the Markov approximation `happens'.

In defining the time-reversed wavefunction we performed a reflection about the origin. We could have instead reversed about a point $t=a$,
\begin{equation}
\psi_{R}(x,a+t) = \psi^{*}(x,a-t).
\end{equation}
It can then be asked whether the equations are symmetric under this transformation. The generalised Langevin equation \eqref{eq:GLE} can straightforwardly be shown to still be symmetric, but the time-symmetric Langevin equation \eqref{eq:qsgnlang} is not, unless $a=0$.
So, the Markov approximation, made at the origin, has preserved time-reversal symmetry \emph{about the origin}, but not about any other point.

Of course, reflecting the system about the time $a$ can be considered as the composition of a \emph{time translation} of the system and bath by an amount $a$, followed by a reflection about the origin.
Thus we see that the reason that the generalised Langevin equations are time-reversal symmetric about an arbitrary temporal point, is that the Hamiltonian \eqref{eq:H} also possesses time translation symmetry.

We are then forced to conclude that it is in fact time translation symmetry that is destroyed by the Markov approximation, in creating the aforementioned temporal boundary distinguishing past from future.
This is consistent with Fig.~\ref{fig:qEnt}, where we can see one can not tell by watching the evolution of the system whether it is moving forwards or backwards in time.
One can only tell whether the system is moving towards or away from the temporal origin.

The question remains as to the significance of the time $t=0$.
This is the point at which the state of the system and bath is uncorrelated.
This is an assumption that is explicitly used to derive Markovian equations of motion.
For example, in section \ref{sec:CL} we assumed that the initial state of the system and bath was uncoupled, with the bath in a thermal state $\rho_{\text{th}}$.
This allowed us to assume that the average of the stochastic force was zero at all times \eqref{eq:zeroavg}.

\section{Reduction to Classical Open Systems}

This paper has focused primarily on derivations of Markovian dynamics in open quantum systems, but the results hold for open classical systems as well.
One can similarly consider a classical degree of freedom coupled to a bath of classical harmonic oscillators, analogous to that considered in section~\ref{redq}.
Indeed the calculation of the equation of motion for the positional degree of freedom proceeds in an identical fashion as in the quantum case, leading to an identical generalised Langevin equation.
The same Markov approximation can then be carried out leading to an analogous Markovian equation of motion as \eqref{eq:qsgnlang} which maintains time-reversal symmetry.
The notable difference is the lack of any zero-point energy effects in the autocorrelation function \eqref{eq:qcorrohm}; and so the classical autocorrelation function takes the form of the high-temperature limit of the quantum case but for all temperatures.

Typically associated with a classical Langevin equation is a Fokker-Planck equation describing the evolution of the corresponding probability distribution.
There are many ways to derive the Fokker-Planck equation \cite{Risken1984,vanKampen2007} but it can be derived straightforwardly by taking the classical limit of it's quantum equivalent, the Brownian motion master equation \eqref{eq:BMsol}.
Let $W(x,p,t)$ be the Wigner function associated to the system degree of freedom of the quantum Brownian particle.
The Brownian motion master equation is then equivalent to the following differential equation for $W$,
\begin{align}
\frac{dW}{dt} = &-\frac{p}{M}\frac{dW}{dx} +\sum_{n=0}^{\infty} \frac{(-\hbar^{2})^{n}V^{(2n+1)}(x)}{(2n+1)!2^{2n}}\frac{d^{2n+1}W}{dp^{2n+1}}\nonumber \\
&+\sgn(t) \gamma \frac{d}{dp} (p W) + \sgn(t) 2\gamma M k_{B} T \frac{d^{2}W}{dp^{2}}. \label{eq:WignerFull}
\end{align}
Taking the classical limit $\hbar \to 0$, we have the equation
\begin{align}
\frac{dW}{dt} = &-\frac{p}{M}\frac{dW}{dx} + \frac{dV}{dx}\frac{dW}{dp}\nonumber \\
&+\sgn(t) \gamma \frac{d}{dp} (p W) + \sgn(t) 2\gamma M k_{B} T \frac{d^{2}W}{dp^{2}}.
\end{align}
If we interpret $W$ as a classical probability density, then this differential equation is precisely a Fokker-Planck equation describing the evolution of a classical Brownian particle under an applied potential $V$.
By performing the time reversal transformations $t \to -t$ and $p \to -p$, the equation can be clearly seen to be time-reversal symmetric, due to the presence of the $\sgn(t)$ factors.
Indeed, the quantum equation \eqref{eq:WignerFull} is also time-reversal symmetric.

\section{Conclusion} \label{concl}

All derivations of the quantum arrow of time must be built upon microscopic dynamics with time-reversal symmetry.
Because of this symmetry, the two directions of time are indistinguishable when considering the microscopic evolution of any many-body system \cite{Lebowitz93}. 
Whether this is the case also for the reduced descriptions of these systems is the core issue addressed in this paper.

In our derivation, two opposing arrows of time are obtained. 
We demonstrate this by analysing several examples of reduced Markovian dynamics resulting from microscopic open quantum systems models, which initially possess time-reversal symmetry. 
In particular we study a system interacting with a bath of harmonic oscillators, which is a standard Hamiltonian model from which to derive the basic equations of dissipative dynamics \cite{Ford65,Cortes85,Caldeira1983}. 
We find that when examined closely the time-reversal symmetry persists into the derived quantum Langevin equation and Brownian motion master equation. 
This is also the case in the Lindblad and Pauli master equations, which are derived from generic time-symmetric Hamiltonians. 

To this end, it is crucial to carefully apply the Markov approximation without implicitly imposing an arrow of time. 
The Markov approximation is typically carried out by choosing a particular distribution of bath oscillator frequencies and coupling constants such that the bath evolves on a much faster time-scale than the system. 
On the time-scale in which the system evolves appreciably, the effects of the interaction with the environment hitherto are washed away. 
The evolution of the system is consequently only determined by its current state and not by its history. 
In these models the time-reversal symmetry is maintained due to memory kernels being even functions of time; so that the time-reversal symmetry persists independently of the particular choice of bath structure which allows the system to be approximated as Markovian. 
As a consequence, the Markov property does not provide an arrow of time in the sense of a violation of time-reversal symmetry, and the system's evolution is Markovian symmetrically from the origin of time.
This persists when these equations are taken to their classical limit
Thus we have support for our proposed extension in Sec. \ref{sec:markov} of the classical definition of Markovianity and of the quantum semigroup property to include temporal symmetry, which was satisfied by all microscopic models analysed in this paper.

Our findings are consistent with the second law of thermodynamics and emphasise the distinction between the concepts of irreversibility and time-reversal symmetry. 
Once the arrow of time and a particular low entropy initial condition at $t=0$ have been chosen, then the von Neumann entropy will increase forward in time from the temporal origin. 
However, a different choice of the arrow of time would have implied the same dynamics.
The Markov approximation applied to the time-reversed evolution leads likewise to the same dissipation and entropy increase. 
Consequently any thermal equilibrium state for a forward-running trajectory is also an equilibrium thermal state for any time-reversed trajectory, and entropy increases in both directions: the system thermalises into both time directions. It is interesting to note that similar conclusions in the classical realm had been previously drawn in \cite{Cohen60} in a statistical mechanics context. Our derivation here is based instead on a fully Hamiltonian picture.

As a result, the quantum arrow of time, descending from the increasing entanglement between a system and its environment \cite{Linden09}, is split into two arrows when the Markov approximation is performed, according to our derivations. 
Increasing entanglement, as well as the decoherence process of the system of interest, follow the time-symmetric quantum master equations here derived, and hence happen symmetrically along opposing time directions. 
That this might have measurable implications in terms of quantum interference between forward and backward processes has been recently proposed in \cite{Rubino21}. 

Furthermore, we speculate that these results may reflect on the cosmological arrow of time. 
In fact, the natural assumption that the universe was dissipative from time zero onwards would suggest that a model of it would rely on the Markov approximation performed at the moment of the Big Bang. 
If so, this would imply that two opposing arrows of time would have emerged from the Big Bang, which would account in turn for the maintenance of time-reversal symmetry despite the ensuing dissipative nature of the universe. We would happen to live in one of them, where dissipation and entropy increase are common experience, but unaware of the existence of the other alternative possibility. 
We notice the striking similarity of this conjecture with the cosmological model proposed in \cite{Barbour14}, where the origin of time is described as the so-called Janus point \cite{Barbour20}. 
Our approach is intrinsically different from the one in \cite{Barbour14}, which is based on considerations of gravitational effects, and, in contrast to \cite{Barbour14}, does not account for the origin of the initial low-entropy state of the universe. 
However, alongside \cite{Barbour14} it leads nonetheless to the similar conclusion that time emerges from $t=0$ in a symmetric fashion, allowing for dynamical evolution in opposing time directions.

It is a fascinating speculation that our analysis of time-reversal symmetry in open systems might impact on current definitions of different arrows of time, and possibly help establish a relationship between them.

\section{Acknowledgements}

The authors would like to thank James Cresser, Cesare Tronci, Paul Bergold, Karim Th{\'e}bault and Giovanni Manfredi for illuminating discussions. 

This work was made possible through the support of the Engineering and Physical Sciences Research Council (EP/W524463/1) and of the John Templeton Foundation (Grant 62210). The opinions expressed in this publication are those of the author(s) and do not necessarily reflect the views of the John Templeton Foundation.

\appendix

\section{Details of the Cumulant Expansion}\label{app:cumulant}
Here, for completeness, we sketch the derivation of the master equation \eqref{eq:masterapprox}, which follows the approach of Gardiner \cite{Gardiner2004}.
This method begins with the adjoint equation
\begin{align}
\dot{\hat{\mu}}(t) = &-\frac{i}{\hbar} \left[\hat{H}_{\textsc{S}},\hat{\mu}(t)\right] + \frac{i}{2\hbar} \left[\set{\gamma\, \sgn(t) \hat{P},\hat{\mu}(t)},\hat{Q}\right] \nonumber \\
&+ \frac{i}{2\hbar} \left[\set{\hat{f}(t),\hat{\mu}(t)},\hat{Q}\right],\label{eq:adjointapp}
\end{align}
from which the master equation can be derived according to the following formula,
\begin{equation}
\hat{\rho}(t) = \expect{\hat{\mu}(t)} := \ptr{\textsc{B}}{\hat{\mu}(t)\hat{\rho}_{\text{th}}},\label{eq:avgapp}
\end{equation}
where $\hat{\rho}_{\text{th}}$ is the thermal state with respect to the Hamiltonian $\hat{H}_{\textsc{b}}$.
The derivation begins by introducing the linear superoperators
\begin{subequations}
\begin{align}
\hat{A}\hat{\mu}(t) = -\frac{i}{\hbar} \left[\hat{H}_{\textsc{s}},\hat{\mu}(t)\right] &+ \frac{i}{2M\hbar} \left[\set{\gamma\, \sgn(t) \hat{P},\hat{\mu}(t)},\hat{Q}\right], \\
\hat{B}\hat{\mu}(t) &= \frac{i}{\hbar}\left[\hat{\mu}(t),\hat{Q}\right],  \\
\hat{\alpha}(t)\hat{\mu}(t) &= \frac{1}{2}\set{\hat{f}(t),\hat{\mu}(t)},
\end{align}
\end{subequations}
and in this notation \eqref{eq:adjointapp} can be written $\dot{\hat{\mu}}(t) = A\mu(t) + B \alpha(t)\mu(t)$.
We then switch to an `interaction picture' by defining
\begin{equation}
\hat{\eta}(t) = \exp(-\hat{A} t)\hat{\mu}(t).
\end{equation}
We can now rewrite \eqref{eq:adjointapp} in more compact notation
\begin{equation}
\dot{\hat{\eta}}(t) = \hat{B}(t)\hat{\alpha}(t)\hat{\eta}(t),
\end{equation}
where $\hat{B}(t) = \exp(-\hat{A} t)\hat{B} \exp(\hat{A} t)$. Note that $\hat{\alpha}(t)$ commutes with itself at different times; and $\hat{\alpha}(t)$ commutes with both $\hat{A}$ and $\hat{B}$, and hence also $\hat{B}(t)$.

We then proceed similarly as in the derivation of the Lindblad equation in section~\ref{sec:Lindblad}.
This differential equation can be formally solved and substituted back into itself. 
We then take the average in the sense of \eqref{eq:avgapp} and the time derivative to get
\begin{align}
\frac{d}{d t} \expect{\hat{\eta}(t)} = \hat{B}(t)&\expect{\hat{\alpha}(t)\hat{\mu}(0)} \\
&+ \int_{0}^{t} \hat{B}(t)\hat{B}(t^{\prime})\expect{\hat{\alpha}(t)\hat{\alpha}(t^{\prime})\hat{\mu}(t^{\prime})}dt^{\prime}. \nonumber 
\end{align}
We now make a number of approximations. In particular, that $\hat{\mu}(t)$ is uncorrelated from $\hat{\alpha}(t)$, so that $\expect{\hat{\alpha}(t)\hat{\mu}(0)} = \expect{\hat{\alpha}(t)\hat{I}}\expect{\hat{\mu}(0)}$, where $\hat{I}$ is the identity operator. Since $\hat{\rho}_{\text{th}}$ is thermal, $\expect{\hat{\alpha}(t)\hat{I}}=0$. We also assume that $\expect{\hat{\alpha}(t)\hat{\alpha}(t^{\prime})\hat{\mu}(t^{\prime})} \approx \expect{\hat{\alpha}(t)\hat{\alpha}(t^{\prime})\hat{I}}\expect{\hat{\mu}(t^{\prime})}$.

Applying these approximations, and converting back to the `non-interacting picture', we have
\begin{equation}
\frac{d\hat{\rho}}{dt}(t) = \hat{A} \hat{\rho}(t) + \int_{0}^{t} \expect{\hat{\alpha}(t)\hat{\alpha}(t^{\prime})\hat{I}} \hat{B} \hat{B}(t^{\prime}-t)\hat{\rho}(t^{\prime}) dt^{\prime}.
\end{equation}
We also assume that $\expect{\hat{\alpha}(t)\hat{\alpha}(t^{\prime})\hat{I}}$ has a short correlation time peaked at $t=t^{\prime}$, allowing us to approximate $\hat{B}(t-t^{\prime})\approx \hat{B}$, and $\hat{\rho}(t^{\prime})\approx\hat{\rho}(t)$; leaving us with the equation
\begin{equation}
\frac{d\hat{\rho}}{dt}(t) = \hat{A} \hat{\rho}(t) - \int_{0}^{t} \expect{\hat{\alpha}(t)\hat{\alpha}(t^{\prime})\hat{I}}dt^{\prime} \frac{1}{\hbar}\left[\left[\hat{\rho}(t),\hat{Q}\right],\hat{Q}\right].
\end{equation}
We note that this master is equation remains time-reversal symmetric at this point and that the integral coefficient $\int_{0}^{t} \expect{\hat{\alpha}(t)\hat{\alpha}(t^{\prime})\hat{I}}dt^{\prime}$ is simply the coefficient $\Gamma(t)$ \eqref{eq:gamma} defined in section \ref{sec:CL}.

Finally, expanding the operator $\hat{A}\hat{\rho}(t)$, we write,
\begin{align}
\dot{\hat{\rho}}(t) = &-\frac{i}{\hbar} \left[\hat{H}_{\textsc{S}},\hat{\rho}(t)\right] + \frac{i}{2\hbar} \left[\set{\gamma\,\sgn(t)  \hat{P},\hat{\rho}(t)},\hat{Q}\right] \nonumber \\
&- \frac{\Gamma(t)}{\hbar^{2}} \left[\left[\hat{\rho}(t),\hat{Q}\right],\hat{Q}\right],
\end{align}
which is precisely \eqref{eq:masterapprox}.

\bibliography{my_references}

\end{document}